\documentclass[conference]{IEEEtran}
\IEEEoverridecommandlockouts
\usepackage{cite}
\usepackage{balance}
\usepackage{changes}
\usepackage{amsmath,amssymb,amsfonts}
\usepackage{algorithmic}
\usepackage{tablefootnote}
\usepackage{graphicx}
\usepackage{multirow}
\usepackage{array}
\usepackage{textcomp}
\usepackage[hidelinks]{hyperref}
\usepackage{url}
\usepackage{changes}
\usepackage{float}
\usepackage[numbers, sort&compress]{natbib}
\usepackage{pifont}
\usepackage[margin=2cm]{geometry} 
\usepackage{caption}

\usepackage{xcolor}
\def\BibTeX{{\rm B\kern-.05em{\sc i\kern-.025em b}\kern-.08em
    T\kern-.1667em\lower.7ex\hbox{E}\kern-.125emX}}

\usepackage{enumitem}
\setlist[itemize]{leftmargin=10pt}
   
\begin{document}

\title{Enhancing IoT Security with CNN and LSTM-Based Intrusion Detection Systems}

\makeatletter % changes the catcode of @ to 11
\newcommand{\linebreakand}{%
  \end{@IEEEauthorhalign}
  \hfill\mbox{}\par
  \mbox{}\hfill\begin{@IEEEauthorhalign}
}
\makeatother % changes the catcode of @ back to 12

\author{\IEEEauthorblockN{Afrah Gueriani}
\IEEEauthorblockA{\textit{LSEA Lab., Faculty of Technology} \\
\textit{University of MEDEA}\\
Medea 26000, Algeria\\
gueriani.afrah@univ-medea.dz }
\and
\IEEEauthorblockN{Hamza Kheddar}
\IEEEauthorblockA{\textit{LSEA Lab., Faculty of Technology} \\
\textit{ University of MEDEA}\\
Medea 26000, Algeria \\
kheddar.hamza@univ-medea.dz}
\and 
\IEEEauthorblockN{Ahmed Cherif Mazari}
\IEEEauthorblockA{\textit{LSEA Lab, Faculty of Science} \\
\textit{ University of MEDEA}\\
Medea 26000, Algeria \\
mazari.ahmedcherif@univ-medea.dz}
}

\makeatletter

\def\ps@headings{%
\def\@oddhead{\parbox[t][\height][t]{\textwidth}{\flushleft

\noindent\makebox[\linewidth]
}
\vspace{0.5cm}
\hfil\hbox{}}%
\def\@oddfoot{\MYfooter}%
\def\@evenfoot{\MYfooter}}
%title page
\def\ps@IEEEtitlepagestyle{%
\def\@evenhead{\scriptsize\thepage \hfil \leftmark\mbox{}}%
\def\@oddfoot{979-8-3503-5026-5/24/\$31.00 \textcopyright 2024   {IEEE} \hfil 
\leftmark\mbox{}}%
\def\@evenfoot{\MYfooter}}
\maketitle

\begin{abstract}
Protecting Internet of things (IoT) devices against cyber attacks is imperative owing to inherent security vulnerabilities. These vulnerabilities can include a spectrum of sophisticated attacks that pose significant damage to both individuals and organizations. Employing robust security measures like intrusion detection systems (IDSs) is essential to solve these problems and protect IoT systems from such attacks. In this context, our proposed IDS model consists on a combination of convolutional neural network (CNN) and long short-term memory (LSTM) deep learning (DL) models. This fusion facilitates the detection and classification of IoT traffic into binary categories, benign and malicious activities by leveraging the spatial feature extraction capabilities of CNN for pattern recognition and the sequential memory retention of LSTM for discerning complex temporal dependencies in achieving enhanced accuracy and efficiency.
In assessing the performance of our proposed model, the authors employed the \textit{new} CICIoT2023 dataset for both training and final testing, while further validating the model's performance through a conclusive testing phase utilizing the CICIDS2017 dataset.
Our proposed model achieves an accuracy rate of 98.42\%, accompanied by a minimal loss of 0.0275. 
False positive rate (FPR) is equally important, reaching 9.17\% with an F1-score of 98.57\%. These results demonstrate the effectiveness of our proposed CNN-LSTM IDS model in fortifying IoT environments against potential cyber threats.
\end{abstract}


\section{Introduction}
\label{sec1}

IoTs have grown notably to sweep the whole world, it involves billions of devices connected to each other without any human interactions (interplay). The IoT generates large data analytics through using sensors, actuators, and control devices. These data are leveraged for diverse tasks and objectives across different fields including healthcare, industry, agriculture, military, and other sectors. The expansive realm of the IoT is proportionate to its exposure to a myriad of threats and cyber attacks that have the potential to compromise the integrity and security of connected devices and networks. Hence, it is imperative to address optimal solutions for countering such behaviors. Moreover, IDSs assume a pivotal role in identifying and mitigating cyber attacks in any network \cite{roopak2019deep, roopak2020intrusion,haggag2020implementing}.

An IDS functions as a monitoring tool, it serves to identify any form of potentially malicious network traffic such as intrusion attempts, viral attacks, and suspicious traffic that pose threats to the security of information systems based on standards when activities deviate from these standards or baseline, the IDS alerts us at an early stage. In the realm of security components, IDSs provide two distinct forms : (1) host-based (HIDS) which focuses on monitoring and analyzing activities transpiring on a server, and (2) network-based (NIDS), tasked with the observation of network activities and communications \cite{albulayhi2021iot}. Numerous organizations opt for a hybrid approach, incorporating both HIDS and NIDS \cite{vinayakumar2019deep}.

Based on the nature of the analysis performed, IDSs are categorized as either signature-based or anomaly-based \cite{garcia2009anomaly}. Signature-based schemes, alternatively referred to as misuse-based, aim to identify predefined patterns, or signatures, within the analyzed data, these systems serve to identify specified and well-known attacks but may fail to detect novel and unfamiliar intrusions \cite{kheddar2023deep,albulayhi2021iot, garcia2009anomaly,elrawy2018intrusion}. Whereas, the anomaly-based IDSs are employed to observe the behavior of a standard network and establish a threshold for detecting deviations from the norm \cite{albulayhi2021iot}. Their main benefit is the ability to detect previously unseen and unknown intrusion activities \cite{kanna2021unified}.

In addressing security threats, IDS software is frequently developed employing artificial intelligence (AI) algorithms, including techniques such as machine learning (ML) and data mining (DM). These methods have proven to be highly effective, in identifying intrusions \cite{kheddar2023deep}. DL is a broader sub-field of ML, its architectural configuration comprises an initial input layer succeeded by a series of hidden layers, which subsequently propagate inputs to the output layer \cite{roopak2019deep}. The CNN represents a DL model extensively applied in different domains, like in \cite{habchi2023ai} for image classifications, in \cite{kheddar2018fourier,djeffal2023noise,kheddar2024deep,kheddar2022high} for speech processing and security, and in \cite{shafiq2020selection} for cyber attacks.
LSTM is a special class of recurrent neural network (RNN), which lies in its capability to be directly applied to raw data without necessitating the usage of any feature selection methods \cite{roopak2019deep}. Nevertheless, LSTM entails a lengthier training duration and demands more computational resources compared to CNN \cite{kanna2021unified}. Hence, this study introduces an advanced unified model, CNN-LSTM, which combines the strength of  CNN and LSTM models.
The below steps constitute the procedural flow and the contribution of this paper.

\begin{itemize}
\item Propose a new IDS using CNN-LSTM hybrid model for enhancing the security of IoT infrastructures,
\item The assessment of the suggested model entails employing a subset of the new CICIoT2023 dataset. The effectiveness and generalizability of the proposed approach are validated through its application to other segments of the same dataset, as well as a distinct CICIDS2017 dataset,
\item Compare our proposed methodology with the existing state-of-the-art work using diverse datasets.
\end{itemize}

The remaining sections of the paper are structured as follows: Section \ref{sec2} gives preliminaries that contain a literature review.  Section \ref{sec3} provides the proposed methodology. This is followed by section \ref{sec4} which delves into the aspects of experimentation, results, and discussion. Section \ref{sec5} covered our conclusion and future directions.

\section{Literature review}
\label{sec2}
DL models have successfully demonstrated considerable efficacy across various fields such as cyber security, image processing, speech recognition, healthcare, and many more. These methods are notably pronounced in their effectiveness compared to traditional ML techniques.

\begin{table*}[h!]
\centering
\caption{Summary of related works on intrusion detection.  The results presented in this table were derived by opting for the most favorable outcome when authors employed various datasets or models.}
\label{tab1}
\scriptsize
\begin{tabular}{m{0.5cm}m{2.6cm}m{6.5cm}m{1.7cm}m{2cm}m{1.7cm}}
\hline
Year & Authors & Focus area & Models & Datasets & Performance (\%)\\
\hline 
2021 & A. Binbusayyis et al. \cite{binbusayyis2021unsupervised} & Unsupervised DL approaches, for network \newline intrusion detection &  CAE-OCSVM & NSL-KDD, UNSW-NB-15  & Acc= 94.28 \\[1mm]

2019 & O. Ibitoye \cite{ibitoye2019analyzing} & Analyzing adversarial attacks against DL \newline for intrusion detection in IoT networks & FNN, SNN & BoT-IoT & Acc=  95.1\\[1mm]

2019 & Chouhan et al \cite{chouhan2019network} & Improving network anomaly detection in IoT networks using a deep learning & CNN & NSL-KDD & Acc= 89.41\\[1mm]

2019 & R. Vinayakumar et al. \cite{vinayakumar2019deep} & Intrusion detection on DL/ML approaches in IoT networks & DNN & KDD Cup 1999, NSL-KDD, UNSW-NB-15, WSN-DS & Acc= 99\\[0.6mm]

\hline
\end{tabular}
\label{tab:related-works}
\end{table*}

The focus area of \cite{binbusayyis2021unsupervised} is the combination of the strengths of convolutional auto-encoders (CAE) and one-class support vector machine learning (SVM), aiming to enhance the performance of network intrusion detection. The proposed work uses the CAE to extract meaningful features from network traffic and detect anomalies, then the authors apply one-class SVM to classify network traffic into normal and abnormal categories. They have used two benchmark datasets to validate their proposed scheme. The results obtained outperform the existing traditional ML and DL.

However in \cite{ibitoye2019analyzing}, the paper investigates adversarial attacks on DL-based IDS for IoT network security, they have employed two DL models including feed-forward neural networks (FNN) and self-normalizing neural networks (SNN) in order to classify intrusion attacks in IoT networks. The results demonstrate that the FNN is well at multi-classification metrics such as Cohen Cappa's score. Upon assessment of adversarial robustness, the SNN demonstrates better resilience when confronted with adversarial samples derived from the IoT dataset. Chouhan et al. \cite{chouhan2019network} introduced an architectural framework termed channel boosted and residual learning-based deep convolutional neural network (CBR-CNN) to address the task of NIDS. The proposed methodology integrates stacked auto-encoders (SAE) and leverages unsupervised training techniques. R. Vinayakumar et al. \cite{vinayakumar2019deep} presents a DNN model that serves to detect and classify unforeseen and unpredictable cyber-attacks in IoT networks in a timely and automatic manner. The performance was tested using many datasets. Table \ref{tab:related-works} presents a collection of various aforementioned studies that have employed IDS based on DL within the framework of IoT environments to detect and mitigate cyber-attacks.

\section{proposed methodology}
This section is reserved for the explanation of our proposed model, which is a combination of CNN-LSTM. This architecture is designed to detect and classify both benign and malicious traffic in a new environment dataset. The proposed scheme has the following steps:

\label{sec3}
\begin{itemize}
\item \textbf{\textit{Data preprocessing:}} Our initial dataset comprised a table encompassing 45 distinct features, consisting of 33 different attack instances along with benign traffic. The features were then extracted and organized into a matrix of vectors to facilitate model training. The subsequent phase involved the conversion of the dataset to two dimensions to align with the input of our model. Additionally, the numerical transformation of the labels into binary form was performed.

\item \textbf{\textit{Data splitting:}} The pre-processed selected sets of the dataset are divided into two segments: The first one is further subdivided into two subsets training and validation, for the second segment, is reserved for the final testing of the model. \\
The initial segment comprises 80\% for training the model and 20\% for validating its performance. Significantly, the second segment pertains to the conclusive testing phase of our model, representing a distinct subset from the initial division. This partition is instrumental in reinforcing the performance evaluation of our model (Figure \ref{fig1}).

\item \textbf{\textit{Model architecture:}} Our proposed model was formulated employing multiple layers, including an input layer, CNN-1D layers, average pooling layers, a flattened layer, and a dense layer as shown in Figure \ref{fig2}, with a combination of two separate DL models, a CNN-1D model and an LSTM model.

First, the model receives a sequence of 45 features, it's the first input layer that takes in raw network traffic data. Then, a series of convolutional layers, batch normalization, and average pooling layers are applied to the input sequence. This is done to extract features and patterns from the sequence. Afterward, dense layers with ReLU activation are used to transform the features into a higher-level representation. The outputs of the last two dense layers are passed through a dense layer with 2 units and a softmax activation function. This is the first task, which is to predict the class of the input sequence. The results of the final two dense layers are also reshaped into a 2D array comprising 16 units, subsequently traversing through multiple LSTM layers, an LSTM layer with a kernel size of 1x256 and a recurrent kernel size of 64x256. This is followed by another LSTM layer with the same configuration. Following this, the outputs of the LSTM layers traverse through a dense layer featuring 2 units and a sigmoid activation function, constituting the second task of predicting the authenticity of malicious traffic. The outcomes of both the classification and prediction tasks are concatenated along the last axis. The concatenated outputs then traverse through a concluding dense layer with 2 units and a softmax activation function, representing the comprehensive output of the model. The output layer has two nodes labeled malicious and benign, indicating that the model is used for binary classification.

\begin{figure}[h!]
\centering
\includegraphics[scale=0.36]{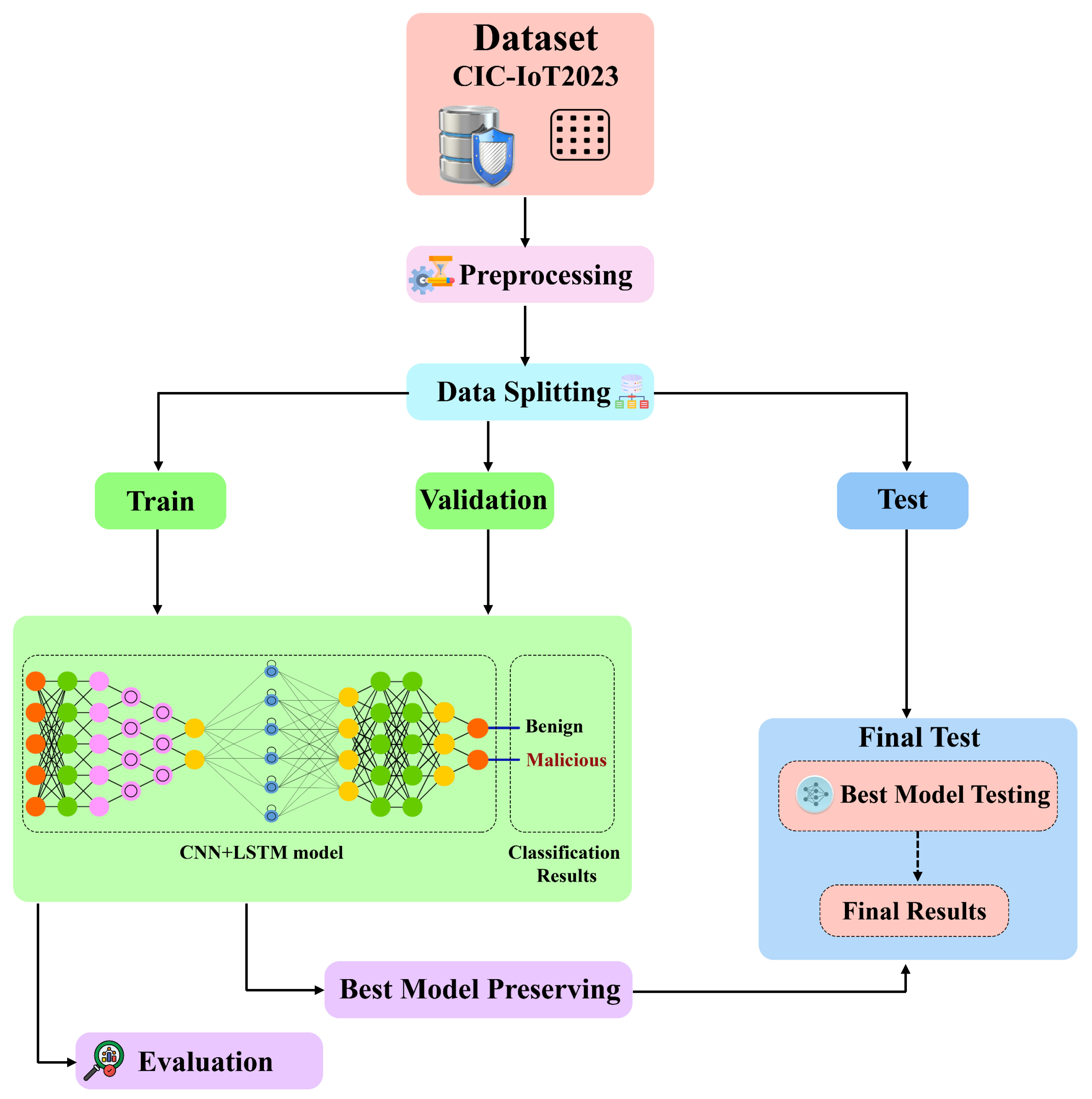}
\caption{The classification framework for CNN-LSTM-based IDS.}
\label{fig1}
\end{figure}

\begin{figure}[h!]
\centering
\includegraphics[scale=1]{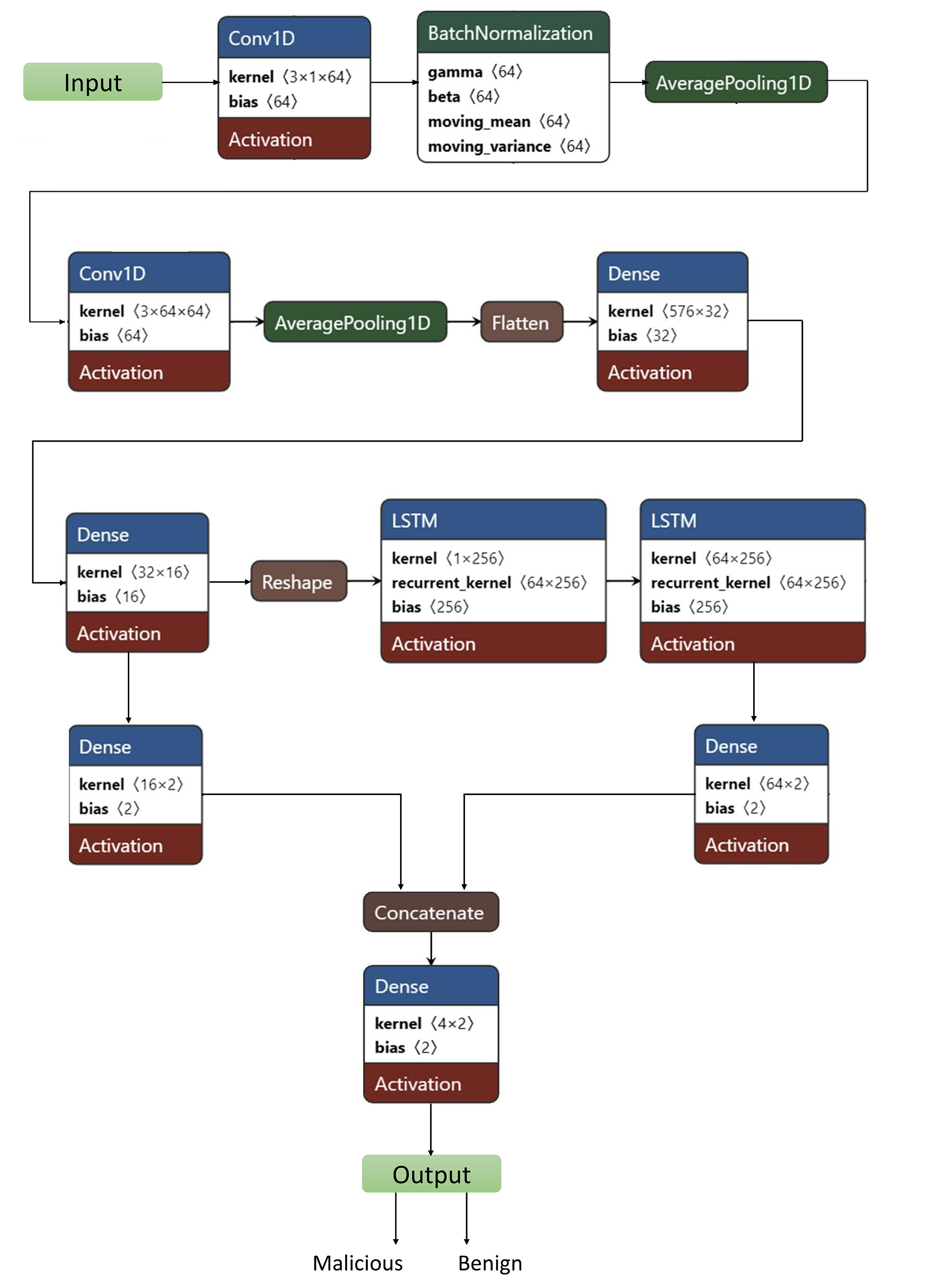}
\caption{CNN-LSTM proposed IDS model.}
\label{fig2}
\end{figure}

\end{itemize}

\section{Experimentation, results and discussion}
\label{sec4}

\subsection{Dataset exploration}

In order to evaluate the performance of our DL-IDS, we have utilized the\textit{ most recent} and \textit{extensive} Canadian institut for cybersecurity 2023 (CIC-IoT2023) dataset\footnote{\url{ https://www.unb.ca/cic/datasets/iotdataset-2023.html}} to carry out the suggested workflow. This dataset boosts the creation of security analytics applications for actual IoTs operations, it contains seven classes with 33 attacks, namely: DDoS, DoS, Recon, Web-based, Brute Force, Spoofing, and Mirai as shown in Table \ref{tab:dataset-attacks}. The Mirai attack system involves a massive DDoS attack targeting IoT devices, both of these categories represent typical and emerging attack classifications within IoT network traffic. Lastly, all attacks are carried out by malicious IoT devices that are directed at other IoT devices. A total of 105 devices were intricately involved in the incident, comprising 67 IoT devices that actively participated in the attacks, while 38 Zigbee and Z-Wave devices were connected to five distinct hubs. This devices includes various categories such as smart home components, cameras, sensors, and microcontrollers. Notably, within this ensemble, certain devices assumed the role of victims, while others assumed an active role as attackers. This dataset encompasses a total of 169 files available in two distinct file formats, namely PCAP and CSV. The CSV files represent processed versions of the PCAP files. It contains almost 47 million instances with both attack and normal data with 45 different features that indicate the different types of attacks. For the hardware computation limit, we have extracted a subset of the dataset for the purpose of our study, comprising approximately 1,191,264 rows representing both attack and normal traffic. 
For the conclusive evaluation of the optimal preservation model, the final test dataset comprises 1,175,692 rows.

\begin{table}[]
\caption{The types of attacks in CIC-IoT2023 dataset}
\label{tab1}
\scriptsize
%\centering
\begin{tabular}{@{}m{1.2cm}m{6.75cm}@{}}
\hline 
Classes & Attacks\\
\hline 
DDoS & ACK fragmentation, UDP flood, SlowLoris, ICMP flood, RSTFIN flood, PSHACK flood,  HTTP flood, UDP fragmentation, TCP flood, SYN flood,  SynonymousIP flood\\[1mm]

Brute force & Dictionary brute force \\[1mm]

Spoofing & ARP spoofing, DNS spoofing \\[1mm]

DoS & TCP flood, HTTP flood, SYN flood, UDP flood \\[1mm]

Recon & Ping sweep, OS scan, Vulnerability scan, Port scan, Host discovery \\[1mm]
Web-based & SQL injection, Command injection, Backdoor malware, Uploading attack, XSS, Browser hijacking \\[1mm]

Mirai & GREIP flood, Greeth flood, UDPPlain \\[1mm]
\hline
\end{tabular}
\label{tab:dataset-attacks}
\end{table}

\subsection{Performance metrics}

The performance of our proposed model for the detection of diverse types of attacks is quantified using standard metrics, including accuracy, precision, recall, F1-score and FPR, which are defined in \cite{gueriani2023deep,kheddar2024deep, kheddarASR2023}. The corresponding equations are presented below:

    \begin{equation}
    \small
        \mathrm{Acc}=\mathrm{\frac{TP+TN}{TP+FP+TN+FN}}
        \label{eq1}
    \end{equation} 

    \begin{equation}
    \small
        \mathrm{Rc}=\mathrm{\frac{TP}{TP+FN}}, \hspace{0.5cm}  \mathrm{Pr}=\mathrm{\frac{TP}{TP+FP}}
        \label{eq2}
    \end{equation}

    \begin{equation}
    \small
        \mathrm{F1-Score}= \mathrm{2 \times \frac{Precision \times Recall}{Precision + Recall}}
        \label{eq6}
    \end{equation}

         \begin{equation}
     \small
        \mathrm{FPR}= \mathrm{\frac{FP}{FP+TN}}
        \label{eq5}
    \end{equation}

Where, the term "true positive" (TP) denotes instances where the IDS accurately identifies an intrusion, while "true negative" (TN) signifies the correct identification of normal traffic. Conversely, "false positives" (FP) denote instances where benign traffic is mistakenly flagged as malicious, and "false negatives" (FN) represent failures of the IDS to detect actual intrusions. A robust F1 score, which integrates precision and recall, is indicative of effective IDS performance, particularly when it reflects low rates of FP and FN \cite{gueriani2023deep}.
\subsection{Results}
This subsection introduces the results of our proposed model. This method used a combination of CNN-LSTM for network security, the results were obtained by splitting the dataset into 80\% for training and 20\% for the validation. In
the conducted experiment, the model underwent training
using the CIC-IoT2023 dataset, encompassing both benign
and malicious network traffic and the training procedure was
executed on Google Colab, employing 25 epochs and the
Adam optimizer.
\begin{itemize}
\item \textbf{\textit{{Accuracy and loss graph:}}} Figure \ref{fig4} shows the accuracy and loss performance of both training and validation established based on the numbers of epochs equal to 25. In Figure \ref{fig4} (b), the model increases as the number of training epochs progresses. This indicates that the model is learning and enhancing its efficiency across successive epochs. The accuracy starts with a value of 98\% and reaches 98.42\% proving the accuracy of the model. On the other hand. In Figure \ref{fig4} (a), the model decreases significantly through the number of epochs, starting approximately from 0.03 to reaching 0.0275 at the last epoch. Evidently, the CNN–LSTM model for the validation demonstrates stability and convergence compared to the training, it indicates that the model is learning effectively and not overfitting the training data.
\end{itemize}

\begin{figure}[h!]
\centering
\includegraphics[height=13cm, width=7.5cm]{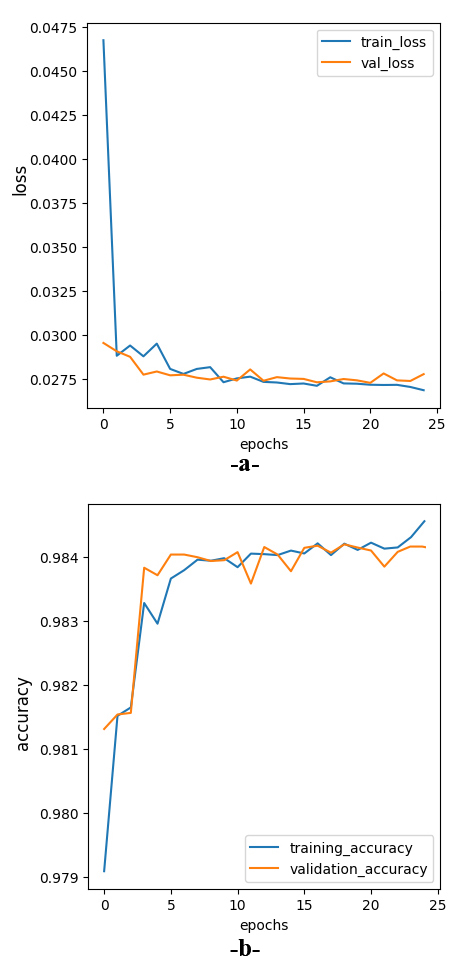}
\caption{Accuracy and loss model during the training phases. (a): train and validation losses. (b) training and validation accuracies.}
\label{fig4}
\end{figure}

\begin{itemize}
\item \textbf{\textit{{Classification Report:}}}
Table \ref{classification-report} presents a detailed evaluation of the binary classification of our system using a set of metrics like precision, recall, F1-score, and support. It is obvious that the model's performance changes through different classes, especially for the first and second classes (normal traffic and attacks).
Concerning the first class, the precision obtained is 90\% compared to other classes, this implies that the model could face difficulties in precisely classifying instances associated with the normal situation category. For the same class, the recall is documented at 61\%, indicating that the model might encounter difficulties in identifying all positive instances (a higher recall indicates fewer false negative results). For this particular class, the F1-score stands at 73\%, which is determined by both precision and recall. 
The challenges encountered may stem from the inherent resemblance between certain features of benign network traffic and malicious attacks. For the other class (attacks), the F1-score reaches a value of 99\%.

It is evident that the model exhibits commendable performance in accuracy, precision, recall, and F1-score. Nevertheless, it is imperative to note that while accuracy provides valuable insights, it alone may not suffice for making the final decision regarding the system's performance. 
\end{itemize}

\begin{itemize}
\item \textbf{\textit{{FPR:}}}
An FPR of 9.17\% is generally considered an acceptable result, signifying that only 9.17\% of instances representing normal traffic were erroneously categorized as attacks. This denotes the classifier's proficiency in accurately discerning the majority of normal traffic instances, a critical factor in mitigating false alarms and enhancing the overall efficacy of the system (Determined using the confusion matrix).

\begin{table}
\caption{Classification report.}
%\centering
\label{tab1}
\begin{tabular}{llllllll}
\hline
&  & Precision & Recall & F1-score & Support \\ [0.6mm]
\hline
& Normal traffic & 90\% & 61\% & 73\% & 8321 \\[0.6mm]

& Attacks & 99\% & 100\% & 99\% & 229932\\ [0.6mm]
\hline

& Accuracy &  &  & 98\% & 238253 \\[0.6mm]

& Macro avg & 95\% & 80\% & 86\% & 238253 \\[0.6mm]

& Weighted avg & 98\% & 98\% & 98\% & 238253 \\[0.6mm]
\hline
\end{tabular}
\label{classification-report}
\end{table}

\item \textbf{\textit{{Confusion matrix:}}}
Referring to Figure \ref{fig5}, it is discerned that the classification performance is notably robust, with an accuracy rate of 90\% for the first class (representing normal traffic), and an even higher accuracy of 99\% for the second class, designated for cyberthreats. Regarding mis-classifications, a marginal 10\% pertains to instances where the first class is erroneously classified as the second, whereas a mere 1\% of attacks are misclassified as normal traffic which substantiates our proposition as delineated in the classification report.

\item \textbf{\textit{{Receiver operating characteristics (ROC):}}} ROC presented in Figure \ref{fig6} indicates the commendable performance of our CNN-LSTM model in the classification of attacks, with high TPR values and low FPR values. The model has an elevated capability to discriminate between normal network traffic and instances of attacks. This is likely because the ROC curve is positioned near the left corner suggesting that the models' predictions are both accurate and precise.

\begin{figure}
\centering
\includegraphics[scale=0.15]{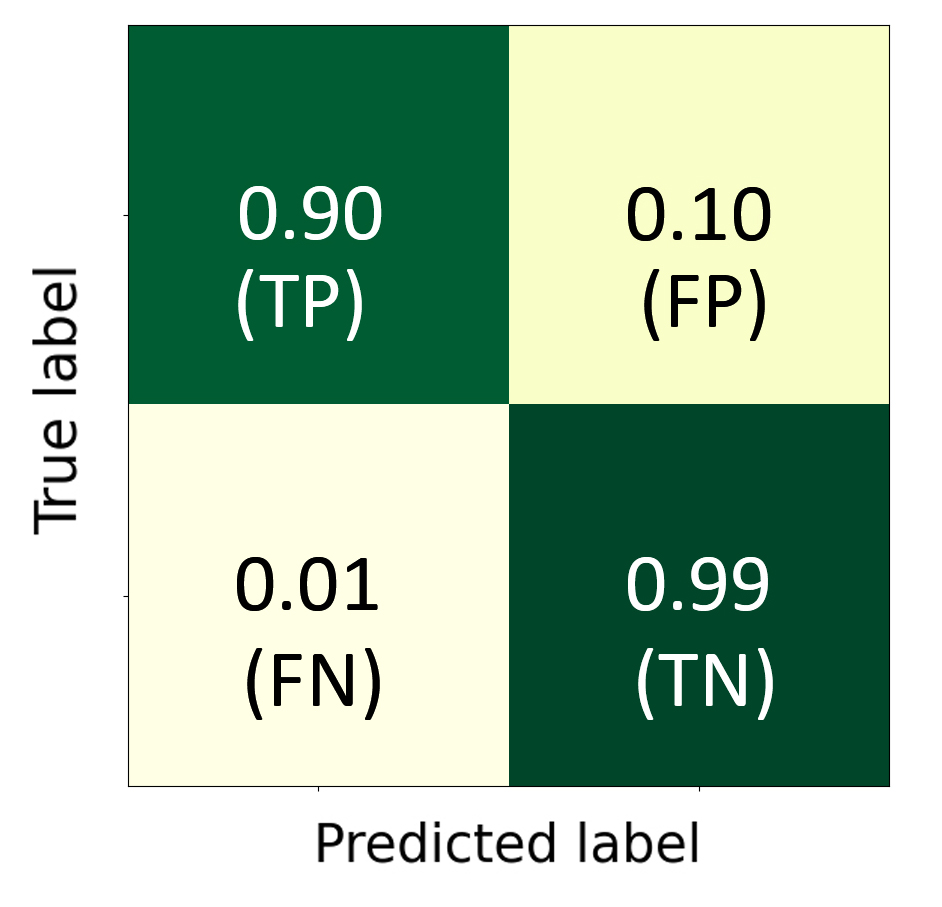}
\caption{Confusion matrix during the training phase.}
\label{fig5}
\end{figure}

\begin{figure}
\centering
\includegraphics[scale=0.55]{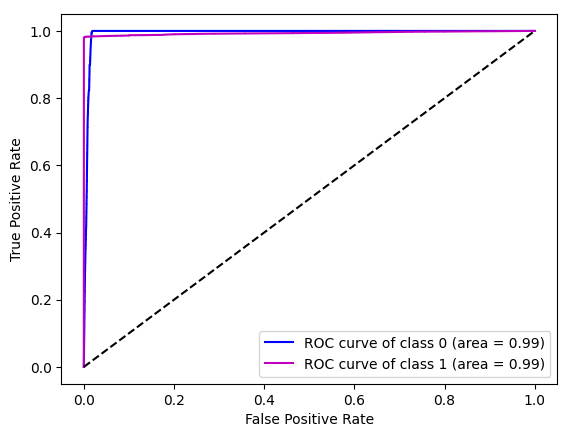}
\caption{ROC curve.}
\label{fig6}
\end{figure}

\item \textbf{\textit{{Generalization verification:}}}
Additional subsets of CICIoT2023 dataset is conducted in our study. The ensuing accuracy from this evaluation attains 98.43\%, accompanied by the precision, recall, and F1-score values of 98.85\%, 98.43\%, and 98.57\%, respectively. Remarkably, the loss metric maintains an analogous value to that observed during the training phase, registering at 0.02\%, while the FPR manifested a numerical value of 9.17\%. It is noteworthy that all obtained results closely align with those of the training model. Besides, we have conducted experiments utilizing an alternative dataset, namely the CICIDS2017. The primary objective is to assess the model's performance across diverse datasets and ascertain its generalization capabilities. The targeted metric for performance evaluation in this context is the same as previous tests, achieving an accuracy rate of 97.45\%, loss of 0.06, precision of 97.17\%, recall 97.15\%, F1-score 97.07\% and FPR 2.08\%. This meticulous examination of the model's proficiency on a distinct dataset serves to reinforce the robustness and reliability of its predictive capabilities, contributing to a more nuanced understanding of its potential applications in real-world scenarios. The confusion matrix of the final test described related to CIC-IoT2023 and CICIDS2017 datasets are in Figure \ref{fig7} (a) and (b) respectively. The results showed similar results to previous tests. However, regarding the confusion matrics of the CICIDS2017,  the classifier correctly identified 98\% instances and misclassified 0.02\% instances as "attacks". Out of instances that are true "attacks", the classifier correctly identified 94\% and misclassified 0.06\% instances as "normal traffic" which shows that the classifier performs well in identifying "normal traffic" but has some error rate in detecting "attacks".

\begin{figure}[]
\centering
\includegraphics[scale=0.15]{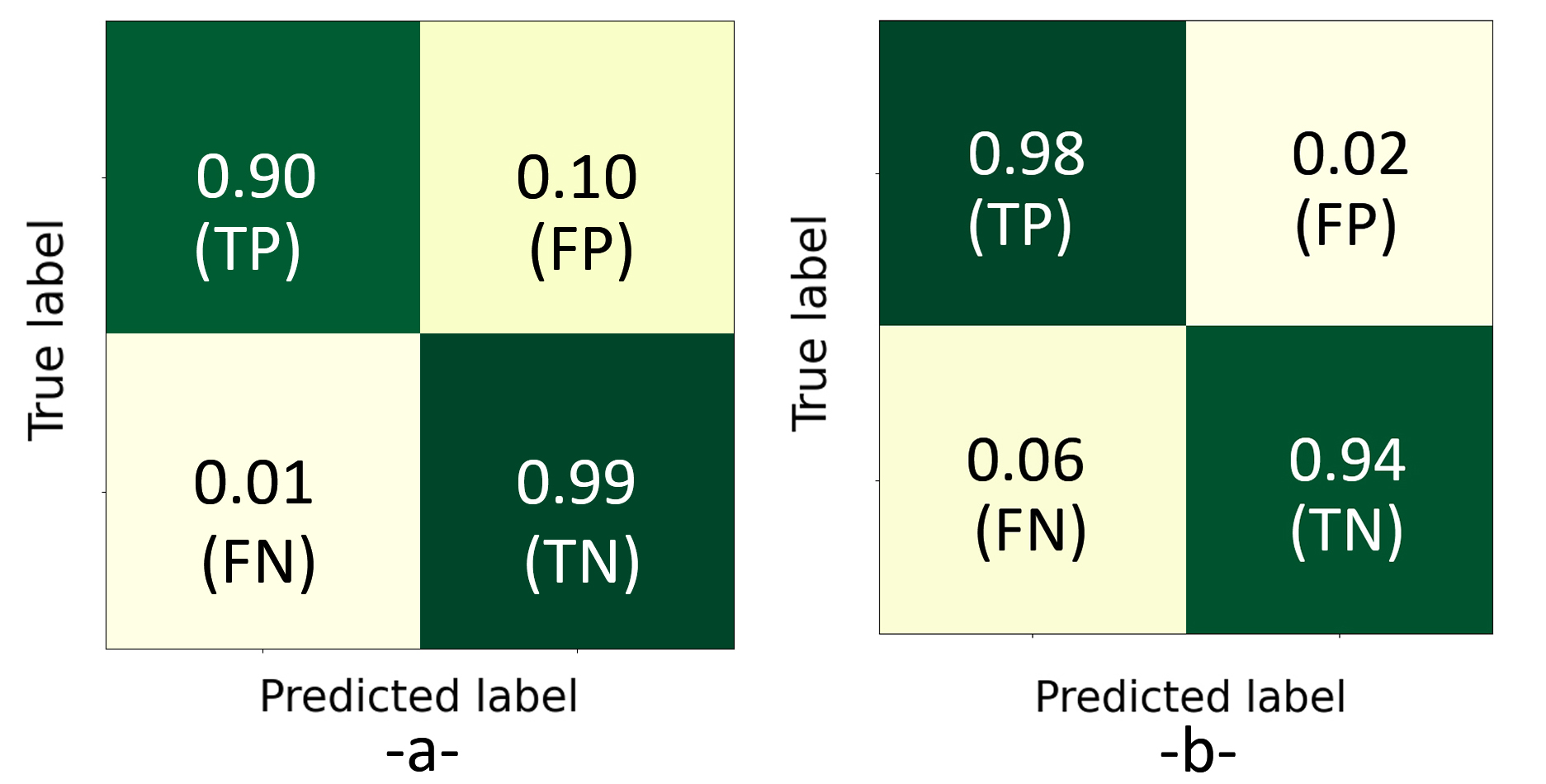}
\caption{Confusion matrix of the generalization verification. (a): the remaining CICIoT2023 subsets. (b): when using the CICIDS2017 dataset.}
\label{fig7}
\end{figure} 

\begin{table*}[h]
\caption{Performance metrics of the proposed model CNN-LSTM compared to state-of-the-art for binary classification.}
\label{tab4}
\scriptsize
\begin{tabular}{m{2.4cm}m{0.6cm}m{1.3cm}m{1.6cm}m{1.4cm}m{0.7cm}m{1.45cm}m{1.2cm}m{1.5cm}m{0.9cm}}
\hline
Work & year & Model & datasets & Accuracy (\%) & Loss & Precision (\%) & Recall (\%) & F1-score (\%) & FPR (\%)\\[0.6mm]
\hline \\
A. Kim et al. \cite{kim2020ai} & 2020 & CNN-LSTM & CICIDS2017 \newline CSIC-2010 & 91, 93 & \ding{55} & 86.47 \newline 98.54 & 94.40 \newline 81.36 & 86.47 \newline 80.65 & \ding{55}\\[0.6mm]

S. S. S. Sugi et al. \cite{sugi2020investigation} & 2020 & LSTM & BoT-IoT & 97.28 & \ding{55} & \ding{55} & \ding{55} & \ding{55} & \ding{55}\\[0.6mm]

M. M. Hassan et al. \cite{hassan2020hybrid} & 2020 & CNN-WDLSTM & UNSW-NB15 & 97.17 & \ding{55} & N: 98, \newline A: 94 & N: 99, \newline A: 82 & N: 98, \newline A: 88 &\ding{55} \\[0.6mm]

W. Yao et al.\cite{yao2023privacy} & 2023 & LSTM-XGboost & CICIoT2023 & 97.7 & \ding{55} & 97.4 & 97.4 & 97.4 & \ding{55}\\[0.6mm]
S. Abba et al. \cite{abbas2024evaluating} & 2024 & RNN & CICIoT2023 & 96.52 & \ding{55} & 96.25 & 96.52 & 96.73 & \ding{55}\\[0.6mm]

Our & 2024 & CNN-LSTM  & CICIoT2023\newline(first subset) & \textbf{98.42} & \textbf{0.0275} & \textbf{98.85} & \textbf{98.42} & \textbf{98.57} & \textbf{9.17}\\ [0.6mm]

 &  &  & CICIoT2023\newline(second subset) & \textbf{98.43} & \textbf{0.0275} & \textbf{98.85} & \textbf{98.43} & \textbf{98.57} & \textbf{9.17}\\ [0.6mm]
 
 &  &  & CICIDS2017 & \textbf{97.46} & \textbf{0.0627} & \textbf{97.17} & \textbf{97.15} & \textbf{97.09} & \textbf{2.08}\\[0.6mm]

\hline
\end{tabular}
\begin{flushleft}
Abbreviations: Normal (N), Abnormal (A)    
\end{flushleft}
\end{table*}

\item \textbf{\textit{{Comparison with state-of-the-art:}}}
Table \ref{tab4} illustrates the outcomes of our system, encompassing a myriad of metrics for comparison with extant works employing different models (CNN-LSTM, LSTM, CNN-WDLSTM, LSTM-XGboost, and RNN) and alternative datasets. It is noteworthy that our study introduces a pioneering dataset. The tabulated data unequivocally demonstrates the superior performance of our proposed model compared to state-of-the-art models across various binary classification datasets, as evidenced by elevated accuracy, lower loss, and heightened recall and precision values.

\end{itemize}

\section{Conclusion}
\label{sec5}
This paper introduces a new IDS that leverages the combined strength of two robust DL models, CNN-LSTM. These models are adept at the detection and binary classification of diverse attacks as well as benign traffic. The training and validation processes of the proposed model are conducted using a specific partition of the recently introduced CICIoT2023 dataset. Subsequently, a distinct subset from this dataset is designated for the conclusive testing phase. In addition to this, to further elucidate the performance of our proposed method, a separate dataset, namely CICIDS2017, is introduced for the final testing evaluation. This methodical approach ensures a comprehensive assessment of the model's generalization across different datasets, thereby enhancing the credibility of its observed performance outcomes. The results obtained demonstrate the efficacy of the CNN-LSTM model in effectively detecting and classifying traffic into binary classification. For future works, we consider using all the CICIoT2023 datasets to achieve more results. Furthermore, integrating a Transformer, such as an attention layer, could significantly enhance the results \cite{kheddar2024automatic,habchi2024machine,djeffal2023automatic}. This layer is adept at capturing intricate features within lengthy dependencies and sequences, thereby refining the overall performance. Given our engagement in binary classification, another next task involves the development of the model to facilitate multi-class classification, encompassing a diverse range of attacks, another prospective avenue to consider in our future work is to study the effectiveness of our proposed approach in a real-time scenario, where the proposed model will be implemented on Raspberry, FPGA, and more.

\section*{Acknowledgment}

The authors acknowledge that the study was partially funded by the PRFU-A25N01UN260120230001 grant from the Algerian Ministry of Higher Education and Scientific Research.

\balance
\bibliographystyle{IEEEtran}
\bibliography{references.bib}

\end{document}